\documentclass[twocolumn,showpacs,preprintnumbers,amsmath,amssymb]{revtex4}

\usepackage{graphicx}% Include figure files
\usepackage{dcolumn}% Align table columns on decimal point
\usepackage{bm}% bold math

\newcommand{\stw}{\mbox{$\sin^2\theta_W$}}
\newcommand{\nubar}[0]{\overline{\nu}}

\newcommand{\nuebar}[0]{\overline{\nu}_{e}}

\newcommand{\nue}[0]{\nu_{e}}

\begin{document}

\title{Can a 3+2 Oscillation Model Explain the NuTeV Electroweak Results?}

\author{J.~S.~Ma }
 \email{jyuko@andrew.cmu.edu}
 \affiliation{Carnegie Mellon University\\Research Experience for 
              Undergraduates at Columbia University}

\author{J.~M.~Conrad, M.~Sorel, G.~P.~Zeller }%
\affiliation{Columbia University}

\date{\today}% It is always \today, today,
             %  but any date may be explicitly specified

\begin{abstract}
  The $\stw$ result from NuTeV falls three standard deviations
  from the value determined by global electroweak fits. It has been suggested
  that one possible explanation for this result could be the 
  oscillation of electron neutrinos in the NuTeV beam to sterile neutrinos. 
  This article examines several cases of masses and mixings for 3+2 neutrino 
  oscillation models which fit the current oscillation data at $99\%$ CL. 
  We conclude that electron to sterile neutrino oscillations can account for 
  only up to a third of a standard deviation between the NuTeV determination 
  of $\stw$ and the standard model.
\end{abstract}

% PACS, the Physics and Astronomy
                             % Classification Scheme.
%\keywords{Suggested keywords}%Use showkeys class option if keyword
                              %display desired
\maketitle

NuTeV precisely determines the electroweak mixing angle through the 
measurement of deep inelastic muon neutrino and antineutrino interactions. 
Although the final value of $\stw$ obtained by this experiment agrees with 
previous neutrino-based measurements, the result is anomalously high when 
compared to the value from global electroweak fits to other data. The NuTeV 
result, $0.2277\pm 0.0013 {\rm (stat)} \pm 0.0009 {\rm (sys)}$~\cite{stw-prl},
is roughly three standard deviations above the standard model value of 
$0.2227\pm0.0004$~\cite{LEPEWWG}. 

Giunti {\em et al.}~\cite{Giunti} have considered neutrino oscillations as 
a possible explanation for the NuTeV results, suggesting that if electron 
neutrinos in the NuTeV beam were oscillating into sterile neutrinos, 
this could effectively lead to the NuTeV observation. Their paper 
demonstrated that a 3+1 (three active and one sterile) neutrino model would 
require very large mixings to the sterile neutrino. Such large mixings are 
now known to be inconsistent with present oscillation limits. In addition,
the proposed oscillations are too large to be consistent with the direct
measurement of the electron neutrino content in the NuTeV beam~\cite{sergey}.
In this paper, we extend the idea in~\cite{Giunti} to oscillation models 
with two sterile neutrinos, {\it i.e.} 3+2 (three active and two sterile) 
neutrino models. For a reveiw of these models and their motivation, see 
Reference~\cite{Sorel}.

%---------------------------------------------------------------------------
\section{The NuTeV Detector and Analysis}
\label{NuTeVIntro}

The design of the NuTeV experiment is described in detail in 
Reference~\cite{nutevNIM}.  This experiment used a high energy 800 GeV proton
beam, taking data in neutrino and antineutrino modes separately. The NuTeV 
detector was located $1450\,m$ downstream from the proton target, and 
consisted of a steel-scintillator target followed by a toroid spectrometer. 
Two types of interactions can occur: charged current events (CC), which 
proceed by $W^{\pm}$ exchange, and neutral current (NC), which proceed by 
$Z^0$ exchange. Both interactions produce a hadron shower of particles in 
the calorimeter. For NC events, the shower is accompanied by an undetectable 
final state neutrino. For CC events, there is instead a muon which can be 
tracked through the calorimeter and the toroid spectrometer. To lowest order
in both QCD and electroweak theory, the ratio of NC to CC rates in neutrino 
and antineutrino scattering relates directly to $\stw$~\cite{lsmith}:
\begin{eqnarray}
R^\nu &=& {{1}\over{2}} - \sin^2 \theta_W + {{5}\over{9}}(1+r)\sin^4 \theta_W 
        \\
R^{\bar \nu} &=& {{1}\over{2}} - \sin^2 \theta_W + 
                 {{5}\over{9}}\left(1+\frac{1}{r}\right)\sin^4 \theta_W,
\end{eqnarray}
where  $R^{\nu,\nubar}$ is the ratio of NC to CC total cross sections and 
$r$ is the ratio of muon neutrino to antineutrino CC total cross sections. 

% Thus,
% \begin{equation}
% {{R^\nu - r R^{\bar \nu}}\over{1-r}} = {{1}\over{2}} - \sin^2 \theta_W.
% \label{PWeq}
% \end{equation}

To extract a value of $\stw$ from the data, NuTeV does not measure total 
cross section ratios, but rather measures experimental ratios of NC to CC 
candidate events. NuTeV differentiates NC and CC interactions simply by the 
measured length of the event~\cite{stw-prl}. NC interactions strictly produce 
hadronic showers and appear as short events in the detector. Longer 
events are likely to be extended by virtue of containing a muon, and thus are 
identified as CC events. The total number of short and long events 
($N^S_{exp}$ and $N^L_{exp}$, respectively) are measured and from them, the 
experimental ratio, $R_{exp} \equiv N^S_{exp}/N^L_{exp}$ is determined in both
the neutrino and antineutrino data. These ratios include the effects of
experimental cuts, cross-talk between candidates in the numerator and
denominator, final state effects, and non-muon neutrino backgrounds.

The second largest background to $N^S_{exp}$, accounting for $\sim5\%$ of 
short events in neutrino mode and $\sim6\%$ in antineutrino mode, 
results from electron neutrino contamination in the beam, the dominant source 
of which are $K^{\pm}_{e3}$ decays. The electron neutrino background is 
determined using beam Monte Carlo tuned to the neutrinos observed from 
$K^{\pm}_{\mu 2}$ decays. This beam prediction is then checked against a 
direct measurement of the electron neutrino content in the NuTeV  
data~\cite{sergey}. The predicted and measured electron neutrinos are found 
to agree: the ratio of measured to Monte Carlo predicted $\nue$ events is 
$1.05 \pm 0.03$ in the neutrino beam and $1.01 \pm 0.04$ in the antineutrino 
beam~\cite{sergey}.

Because of their event topology, electron neutrino interactions all appear
as short events in the NuTeV detector. Their contribution (in addition to
other corrections which are not shown here) is explicitly included as a 
modification to the number of short events appearing in the numerator of 
the predicted experimental ratio:
\begin{equation}
        R_{exp}^{MC} = (N_{exp}^{S,MC} + N_{exp}^{\nue,MC})/
                        N_{exp}^{L,MC}
\end{equation}
in both neutrino and antineutrino modes. Any overestimate of the electron 
neutrino contribution (for example, that would result from neglecting 
$\nue \rightarrow \nu_s$ oscillations) would lead to an overestimate of the 
predicted ratio, $R_{exp}^{MC}$, and hence a larger measured value of NuTeV
$\stw$. Importantly, any adjustment to the electron neutrino flux needed
to reduce the NuTeV $\stw$ value and bring the result into better agreement 
with expectation must additionally satisfy the direct constraint from the 
NuTeV data itself~\cite{sergey}. Here, we consider several such possibilities.

%---------------------------------------------------------------------------
\section{Neutrino Oscillations}
\label{NuTeVosc}

%If the electron neutrinos in the beam were to oscillate to sterile neutrinos, 
%then $N^e_{MC}$ would be an overestimate of the number of electron neutrino 
%interactions in the detector.  

To gain an understanding of the potential impact of electron neutrino 
oscillations in NuTeV, first consider the approximation of a simple 
two-flavor ($\nue \rightarrow \nu_s$) oscillation probability:
\begin{equation}
   P = \sin^2 2\theta \sin^2 (1.27 \Delta m^2 L/E).
\label{2flav}
\end{equation}
The fundamental parameters describing the oscillation are $\sin^2 2\theta$, 
the mixing between the flavors, and $\Delta m^2$, the squared mass difference 
between the neutrinos. The experimental parameters are $L$, the baseline, and 
$E$, the incident neutrino energy. In general, oscillations become observable 
when $\Delta m^2 L/E \sim 1$ or larger.  The high beam energy and short 
baseline of NuTeV lead to a small value of $L/E$, therefore requiring a large 
value of $\Delta m^2$ (of a few eV$^2$ or greater) to compensate.

In the few eV$^2$ range of $\Delta m^2$, the Bugey reactor experiment sets the
best limit on the mixing angle for $\overline{\nue}$ 
disappearance~\cite{Bugey}. Reference~\cite{Giunti} shows that, in a 3+1 
model, the NuTeV result implies $\Delta m^2 \sim 10$ eV$^2$ and 
$\sim \sin^2 2\theta = 0.4$, which would have produced a clear signal in 
Bugey; therefore, this solution is directly excluded. The situation 
for 3+2 models, however, could differ. In such a model, the probability for 
electron neutrino disappearance at high $\Delta m^2$, with mixing matrix 
parameters denoted as $U$ rather than mixing angles, is given by:
\begin{eqnarray}
  P=& 4[U_{e4}^2(1-U_{e4}^2)\sin^2 x_{41} + U_{e5}^2(1-U_{e5}^2)\sin^2 
    x_{51} \nonumber   \\ 
    & - U_{e4}^2 U_{e5}^2 (\sin^2 x_{41} + \sin^2 x_{51} - \sin^2 x_{54})] & 
\end{eqnarray}
where $x_{ij} = 1.27 \cdot \Delta m^2_{ij}(L/E)$. Such 3+2 models can fit the 
world's oscillation data with many different values of mass splittings and 
mixing parameters~\cite{Sorel}.  Those which provide a good description of
the data tend to have $\Delta m_{14}^2 \sim 1$ eV$^2$ and 
$\Delta m_{51}^2 > 10$ eV$^2$.   

\begin{figure}
\vspace{5mm}
\centering
\includegraphics[width=7.0cm, clip=true]{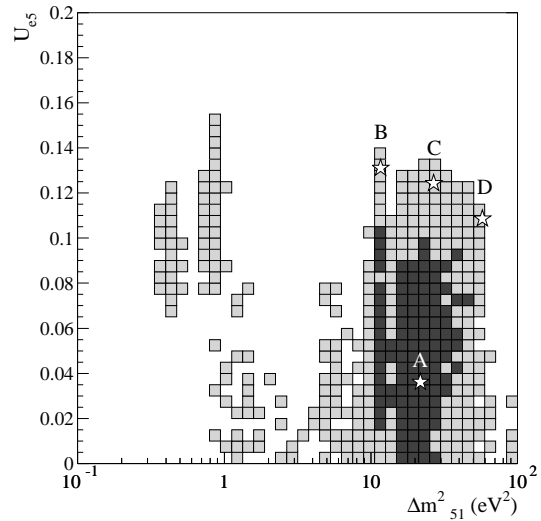}
  \vspace{-0.1in}
  \caption{90\% (dark grey) and 99\% (light grey) CL allowed regions in 
           $(\Delta m^2_{51}, U_{e5})$-space for CP-conserving (3+2) models.  
           The stars labeled A, B, C, and D indicate the four models 
           evaluated for their impact on the NuTeV $\stw$ analysis.}
\label{allowed}
\end{figure}

For NuTeV, which has a small $L/E$, in the case where $\Delta m^2_{41} << 10$
eV$^2$, (which is the 3+2 best fit case), and taking $x_{41}=0$ and
$x_{54}=x_{51}$, the oscillation probability simplifies to:
\begin{equation}
\label{eqn:ue5}
  P = 4U_{e5}^2(1-U_{e5}^2)\sin^2 x_{51},
\end{equation}
analogous to Equation~\ref{2flav}. Figure~\ref{allowed} shows the
$90\%$ and $99\%$ CL allowed regions in the $\Delta m_{51}^2$ and $U_{e5}$ 
mixing space, as determined using the methods described in 
Reference~\cite{Sorel}. The short-baseline data yielding the parameter space
relevant for this study include Bugey~\cite{Bugey}, CCFR84~\cite{CCFR84}, 
CDHS~\cite{CDHS}, CHOOZ~\cite{CHOOZ}, KARMEN2~\cite{KARMEN2}, 
LSND~\cite{LSND}, and NOMAD~\cite{NOMAD}.

%---------------------------------------------------------------------------
\section{Impact of 3+2 Models on the NuTeV Electroweak Results}
\label{NuTeVosc}

\begin{table}[tb]
\begin{center}
{
\begin{tabular}{|c|c|c|c|} \hline
no oscillation    &  NuTeV measurement & expectation  & deviation     \\ \hline
$R_{exp}^\nu$        & $0.3916\pm 0.0013$ & 0.3950  & $-2.6 \,\sigma$ 
\\ \hline
$R_{exp}^{\bar \nu}$ & $0.4050\pm 0.0028$ & 0.4066  & $-0.6 \,\sigma$ 
\\ \hline
$\stw$               & $0.2277\pm 0.0016$ & 0.2227  & $+3.0 \,\sigma$ 
\\ \hline \hline
model B          &  NuTeV measurement & expectation  & deviation     \\ \hline
$R_{exp}^\nu$        & $0.3916\pm 0.0013$ & 0.3949  & $-2.5 \,\sigma$ 
\\ \hline
$R_{exp}^{\bar \nu}$ & $0.4050\pm 0.0028$ & 0.4065  & $-0.5 \,\sigma$ 
\\ \hline
$\stw$               & $0.2277\pm 0.0016$ & 0.2227  & $+3.0 \,\sigma$ 
\\ \hline \hline
model C          &  NuTeV measurement & expectation  & deviation     \\ \hline
$R_{exp}^\nu$        & $0.3916\pm 0.0013$ & 0.3948  & $-2.5 \,\sigma$ 
\\ \hline
$R_{exp}^{\bar \nu}$ & $0.4050\pm 0.0028$ & 0.4062  & $-0.4 \,\sigma$ 
\\ \hline
$\stw$               & $0.2275\pm 0.0016$ & 0.2227  & $+2.9 \,\sigma$ 
\\ \hline \hline
model D          &  NuTeV measurement & expectation  & deviation     \\ \hline
$R_{exp}^\nu$        & $0.3916\pm 0.0013$ & 0.3945  & $-2.2 \,\sigma$ 
\\ \hline
$R_{exp}^{\bar \nu}$ & $0.4050\pm 0.0028$ & 0.4059  & $-0.3 \,\sigma$ 
\\ \hline
$\stw$               & $0.2271\pm 0.0016$ & 0.2227  & $+2.7 \,\sigma$ 
\\ \hline 
\end{tabular}}
\caption{Comparison of NuTeV electroweak results assuming no 
         $\nue$ oscillations (default) to the results assuming three 
         CP-conserving 3+2 oscillation models (B,C,D in Fig.~\ref{allowed}). 
         In all cases, the same event selection criteria as 
         in~\cite{stw-prl} is applied.}
\label{tab:errsig}
\end{center}
\end{table}

Given the increase in possible parameter space inherent in 3+2 models, we
ask whether there is an oscillated $\nue$ flux which can account for the 
NuTeV electroweak results. To answer this question and demonstrate the impact 
of a 3+2 model neutrino oscillations on the NuTeV data, four representative 
points are selected within the allowed region as indicated in 
Figure~\ref{allowed}. These points are the best fit model (A), a high mixing 
model with lower mass (B), a high mixing model with higher mass (C), and a 
high mass model (D). The best fit point (A) was eventually omitted from the 
study because it represented such a small correction to the unoscillated flux 
(as expected from Equation~\ref{eqn:ue5}) that it had a negligible impact on 
NuTeV $\stw$.

\begin{figure}
\vspace{5mm}
\centering
\includegraphics[width=7.4cm,bb=21 175 574 627,clip=true]{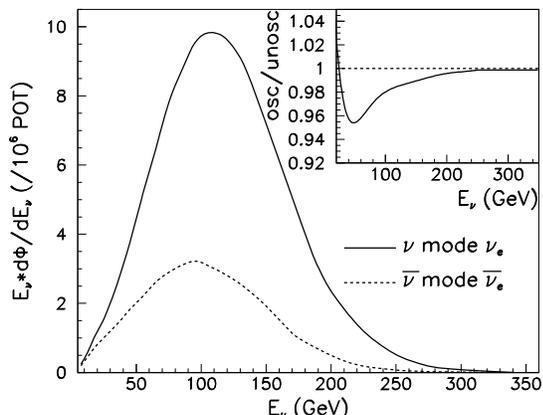}
\vspace{-0.3in}
\caption{Unoscillated NuTeV $\nue$ flux prediction in each mode. Inlay 
         shows the ratio of the predicted oscillated/unoscillated $\nue$ 
         fluxes 
         as a function of energy for Model D. The change is within the 
         errors of the NuTeV $\nue$ measurement~\cite{sergey}.}
\label{example}
\end{figure}

For each set of possible parameters, a $\nue$ survival probability is 
calculated as a function of energy. This probability is then used to correct 
the estimated NuTeV $\nue$ and $\nuebar$ fluxes. The resulting 
difference is very small in all cases, being largest for the high mass Model D
(Figure~\ref{example}). The total integrated $\nue$ flux prediction changes
by $0.2\%$ (Model B), $0.8\%$ (Model C), and $1.8\%$ (Model D), hence
satisfying the NuTeV $\nu_e$ data constraint~\cite{sergey}. Based on each 
of the ``oscillated'' $\nue$ and $\nuebar$ fluxes, the $R_{exp}^{MC}$ 
predictions are then recalculated in both neutrino and antineutrino modes, 
and a new value of $\stw$ is extracted. In Table~\ref{tab:errsig}, we report 
the deviation of the measured $R_{exp}^{\nu,\nubar}$ and $\stw$ values for 
the three models: B, C, and D. The magnitude and 
sign of the shift indicates the expectation if the NuTeV data had been 
analyzed using the oscillated electron neutrino fluxes. All of the 
adjusted fluxes move the NuTeV results into better agreement with the 
standard model, by construction. Model D provides the largest 
impact: shifting $R_{exp}^\nu$ and $R_{exp}^{\nubar}$ into better agreement 
with expectation by $0.4 \, \sigma$ and $0.3 \,\sigma$, respectively. Model 
D reduces the NuTeV $\stw$ discrepancy with the standard model from 
$3.0 \,\sigma$ to $2.7 \,\sigma$.

%---------------------------------------------------------------------------
%\section{Conclusions}
%\label{results}

\vspace{0.2in}
We have studied three points representative of extreme masses and 
mixings within the allowed region of 3+2 models, and conclude that 
$\nue \rightarrow \nu_s$ oscillations in this model do not yield a 
significant impact on NuTeV's electroweak results.  The largest shift in 
$\sin^2 \theta_W$ is created by a high mass model (e.g. model D), but even 
such a high mass model would only affect the NuTeV value of $\stw$ 
by roughly $0.3 \,\sigma$.  Therefore, a 3+2 model with 
$\nue \rightarrow \nu_s$ oscillations cannot explain the NuTeV electroweak
results by itself.

%---------------------------------------------------------------------------
\subsection*{Acknowledgments}

We thank the NSF for support of this work through the REU grant:
NSF-PHY-01-39145. We also thank S.~Avvakumov, K.~McFarland, and M.~Shaevitz 
for useful discussion as well as the rest of the NuTeV collaboration.

%------------------------------- bibliography ---------------------------
\bibliography{nutevstudynew}

%------------------------------------------------------------------------
\end{document}